# Enhancing Graph Attention Neural Network Performance for Marijuana Consumption Classification through Large-scale Augmented Granger Causality (lsAGC) Analysis of Functional MR Images

Ali Vosoughi[a], Akhil Kasturi[a], and Axel Wismüller[a,b,c,d]

[a]Department of Electrical and Computer Engineering, University of Rochester, NY, USA

[b]Department of Imaging Sciences, University of Rochester, NY, USA [c]Department of Biomedical Engineering, University of Rochester, NY, USA [d]Faculty of Medicine and Institute of Clinical Radiology, Ludwig Maximilian University, Munich, Germany

## ABSTRACT

In the present research, the effectiveness of large-scale Augmented Granger Causality (lsAGC) as a tool for gauging brain network connectivity was examined to differentiate between marijuana users and typical controls by utilizing resting-state functional Magnetic Resonance Imaging (fMRI). The relationship between marijuana consumption and alterations in brain network connectivity is a recognized fact in scientific literature. This study probes how lsAGC can accurately discern these changes. The technique used integrates dimension reduction with the augmentation of source time-series in a model that predicts time-series, which helps in estimating the directed causal relationships among fMRI time-series. As a multivariate approach, lsAGC uncovers the connection of the inherent dynamic system while considering all other time-series. A dataset of 60 adults with an ADHD diagnosis during childhood, drawn from the Addiction Connectome Preprocessed Initiative (ACPI), was used in the study. The brain connections assessed by lsAGC were utilized as classification attributes. A Graph Attention Neural Network (GAT) was chosen to carry out the classification task, particularly for its ability to harness graph-based data and recognize intricate interactions between brain regions, making it appropriate for fMRI-based brain connectivity data. The performance was analyzed using a five-fold cross-validation system. The average accuracy achieved by the correlation coefficient method was roughly 52.98%, with a 1.65 standard deviation, whereas the lsAGC approach yielded an average accuracy of 61.47%, with a standard deviation of 1.44. A random guess method yielded an average accuracy of about 47.05%, with a standard deviation of around 6.25. The study indicates that lsAGC, when paired with a Graph Attention Neural Network, has the potential to serve as a novel biomarker for pinpointing marijuana users, offering a superior and consistent classification strategy over traditional functional connectivity techniques, including the random guess method. The suggested method enhances the body of knowledge in the field of neuroimaging-based classification and emphasizes the necessity to consider directed causal connections in brain network connectivity analysis when studying marijuana's effects on the brain.

## 1  INTRODUCTION

Marijuana's longstanding status as an illegal drug, coupled with the apprehensions regarding its negative influence on brain processes and cognitive abilities, has spurred the quest for precise biomarkers to evaluate its ramifications. Studies involving resting-state functional MRI (rs-fMRI) have unveiled that consistent marijuana consumption may modify the brain's connectivity [1,2]. Techniques known as Multi-Voxel Pattern Analysis (MVPA) have been utilized to derive biomarkers from rs-fMRI data, which helps in distinguishing the connectivity patterns between individuals with neurological complications and healthy subjects [1,3].

Most MVPA studies have historically leveraged cross-correlation to assemble a functional connectivity profile, offering optimistic results in terms of information differentiation extracted from fMRI data. However, this method does not encompass directed connectivity, and thus, vital information in the fMRI data might be overlooked. In our research, we tackle this issue by presenting large-scale Augmented Granger Causality (lsAGC) as an innovative technique for estimating the directed causal links within fMRI time-series [4]. lsAGC employs dimension reduction and predictive modeling of time-series to effectively deduce directed causal dependencies in broad scenarios.

We explore whether variations in directed connectivity might function as biomarkers for regular marijuana intake in adults who were diagnosed with ADHD in childhood, and whether these directed evaluations surpass traditional functional connectivity metrics in classifying frequent marijuana users from typical controls. To this end, we incorporate lsAGC within the MVPA framework to infer directed causal connections among fMRI time-series. For enhancing the classification performance, we engage a Graph Attention Neural Network (GAT) [5] for classification, as it excels in processing graph-related data and discerning complex connections between brain areas, rendering it apt for our brain connectivity data obtained from fMRI.

This paper details the outcomes of our classification trials using lsAGC and GAT and juxtaposes them with traditional cross-correlation-centered approaches. Our discoveries underline lsAGC's merit as a pertinent biomarker for frequent marijuana use and accentuate the efficacy of employing GAT in this setting for elevated classification performance.

This work is embedded in our group's endeavor to expedite artificial intelligence in biomedical imaging by means of advanced pattern recognition and machine learning methods for computational radiology and radiomics, *e.g.*, [6–84].

## 2    DATA

The ACPI dataset's resting-state fMRI scan parameters can be found on the project's website [85]. We utilized preprocessed data from the MTA 1 dataset [85] of the ACPI database. The dataset initially included 126 subjects, consisting of 101 males and 25 females, with ages between 21-27 years. Of these, 86 were diagnosed with ADHD (68%), and 62 of them regularly used marijuana (49%). Preprocessing of the raw 4D rs-fMRI data was performed using a Configurable

Pipeline for the Analysis of Connectomes (C-PAC) [86] and the Advanced Normalization Tools (ANTs) pipeline. This preprocessing involved the removal of the first five fMRI volumes, anatomical registration, tissue segmentation, functional registration in the Montreal Neurological Institute (MNI) space, functional masking, temporal bandpass filtering (0.01 - 0.1 Hz), motion correction, spatial smoothing, and various nuisance corrections [87]. The MODL parcellation, which is a precomputed atlas for regions definition built using a form of online dictionary learning, was used [1]. For our study, we utilized a subset of the dataset, consisting of 60 subjects, with 32 healthy individuals and 28 diagnosed with marijuana consumption.

Table 1: Distribution of subjects in the subset dataset are listed in this table.

| Subject Group | Total Subjects | Marijuana Users |
|---|---|---|
| Healthy Controls | 32 | No |
| Marijuana Users | 28 | Yes |

## 3 METHODS

### 3.1 Large-scale Augmented Granger Causality (lsAGC)

Large-scale Augmented Granger Causality (lsAGC) has been developed based on 1) the principle of original Granger causality, which quantifies the causal influence of time-series $\mathbf{x_s}$ on time-series $\mathbf{x_t}$ by quantifying the amount of improvement in the prediction of $\mathbf{x_t}$ in presence of $\mathbf{x_s}$. 2) the idea of dimension reduction, which resolves the problem of the tackling a under-determined system, which is frequently faced in fMRI analysis, since the number of acquired temporal samples usually is not sufficient for estimating the model parameters [88].

Consider the ensemble of time-series $X \in \mathsf{R}^{N \times T}$, where $N$ is the number of time-series (Regions Of Interest – ROIs) and $T$ the number of temporal samples. Let $X = (\mathbf{x_1}, \mathbf{x_2}, ..., \mathbf{x_N})^T$ be the whole multidimensional system and $x_i \in \mathsf{R}^{1 \times T}$ a single time-series with $i = 1, 2, ..., N$, where $\mathbf{x_i} = (x_i(1), x_i(2), ..., x_i(T))$. In order to overcome the under-determined problem, first X will be decomposed into its first $p$ high-variance principal components $Z \in R^{p \times T}$ using Principal Component Analysis (PCA), i.e.,

$$Z = WX, \qquad (1)$$

where $W \in R^{p \times N}$ represents the PCA coefficient matrix. Subsequently, the dimension-reduced time-series ensemble Z is augmented by one original time-series $\mathbf{x_s}$ yielding a dimension-reduced augmented time-series ensemble $Y \in R^{(p+1) \times T}$ for estimating the influence of $\mathbf{x_s}$ on all other time-series.

Following this, we locally predict the dimension-reduced representation Z of the original highdimensional system X at each time sample $t$, i.e. $Z(t) \in R^{p \times 1}$ by calculating an estimate $\hat{Z}_{\mathbf{x_s}}(t)$. To this end, we fit an affine model based on a vector of $m$ vector of m time samples of $Y(\tau) \in R^{(p+1) \times 1} (\tau = t - 1, t - 2,...,t - m)$, which is $\mathbf{y}(t) \in R^{m \cdot (p+1) \times 1}$, and a parameter matrix $A \in R^{p \times m \cdot (p+1)}$ and a constant bias vector $\mathbf{b} \in R^{p \times 1}$,

$$\hat{Z}_{\mathbf{x_s}}(t) = A\mathbf{y}(t) + \mathbf{b}, \, t = m + 1, m + 2,...,T. \qquad (2)$$

Subsequently, we use the prediction $\hat{Z}_{\mathbf{x_s}}(t)$ to calculate an estimate of X at time $t$, i.e. $X(t) \in R^{N \times 1}$ by inverting the PCA of equation (2), i.e.

$$X = W^{\dagger}Z, \qquad (3)$$

where $W^{\dagger} \in R^{N \times p}$ represents the inverse of the PCA coefficient matrix $W$, which is calculated as the *Moore Penrose* pseudoinverse of $W$. Now $\hat{X}_{\setminus \mathbf{x_s}}(t)$, which is the prediction of $X(t)$ without the information of $\mathbf{x_s}$, will be estimated. The estimation process is identical to the previous one, with the only difference being that we have to remove the augmented time-series $\mathbf{x_s}$ and its corresponding column in the PCA coefficient matrix $W$.

The computation of a lsAGC index is based on comparing the variance of the prediction errors obtained with and without consideration of $\mathbf{x_s}$. The lsAGC index $f_{\mathbf{x_s} \rightarrow \mathbf{x_t}}$, which indicates the influence of $\mathbf{x_s}$ on $\mathbf{x_t}$, can be calculated by the following equation:

$$f_{\mathbf{x_s} \rightarrow \mathbf{x_t}} = \log \frac{\text{var}(e_s)}{\text{var}(e_{\setminus s})}, \qquad (4)$$

where $e_{\setminus s}$ is the error in predicting $\mathbf{x_t}$ when $\mathbf{x_s}$ was not considered, and $e_s$ is the error, when $\mathbf{x_s}$ was used.

### 3.2 Graph Attention Network (GAT)

Graph Attention Networks (GATs) have emerged as a vital advancement in graph-based deep learning [5], marking a significant departure from conventional Graph Convolutional Networks (GCNs) [89]. Where traditional GCNs treat all neighboring nodes equally, GATs utilize attention mechanisms to assign different importance levels to different nodes, enabling the capture of more complex patterns in data. This attribute is crucial in recognizing and modeling intricate relationships in graph-structured data, an aspect that conventional models often fall short of.

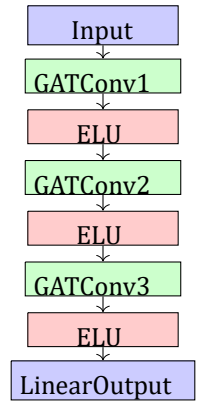

Figure 1: Architecture of the proposed lsAGC + GAT model for classification based on the concept of Granger causality and graph attention neural networks. In this figure, we demonstrate how the GAT model successively refines the graph's representation. Beginning with the initial feature transformation and followed by multiple attention-guided convolutions and ELU activations, the model captures the intricate structure and relationships encoded within the adjacency matrix. The final linear transformation delivers the predictions, effectively leveraging the graph's topology to inform the underlying task. This layer-by-layer breakdown illustrates the coherent integration of the graph's structure and node features, achieved through the thoughtful design of the GAT architecture.

Combination of lsAGC and GATs offer an elegant and robust means to model complex relationships in graph-structured data, surpassing traditional methods in both flexibility and expressive power. The architecture of the pipeline of the lsAGC + GAT model is depicted in Figure 1. In the following subsections we detail the method.

### 3.2.1 Adjacency Matrix

Given a directed graph $G = (V,E)$ with $N$ nodes, the adjacency matrix $\mathbf{F} \in \mathbb{R}^{N \times N}$ is defined such that $F_{ij} = 1$ if there is an edge between nodes $i$ and $j$, and $F_{ij} = 0$ otherwise. Our method, lsAGC, employs the measure $f_{\mathbf{x_s} \to -\mathbf{x_t}}$ to determine whether node $i$ (representing $x_s$) has a causal influence on node $j$ (representing $x_t$).

### 3.2.2 Attention Mechanism

Graph Convolutional Networks (GCNs) inherently treat all neighbors equally, often leading to limitations in expressiveness. This poses challenges in the application of GCNs to directional graphs, which are prevalent in our method, the lsAGC. In contrast, Graph Attention Networks (GATs) employ an attention mechanism that emphasizes the more relevant parts of the graph, facilitating a nuanced understanding of the data's underlying structure.

The attention mechanism in GATs is implemented by calculating attention scores between nodes through a series of learnable parameters and non-linear transformations. The attention scores between nodes $i$ and $j$ are computed as follows:

$$e_{ij} = \text{LeakyReLU}\left(\mathbf{a}^T [\mathbf{W}\mathbf{h}_i \| \mathbf{W}\mathbf{h}_j]\right), \tag{5}$$

$$\alpha_{ij} = \text{softmax}_j(e_{ij}) = \frac{\exp(e_{ij})}{\sum_{k \in \mathcal{N}(i)} \exp(e_{ik})}. \tag{6}$$

Here, $\mathbf{h}_i$ denotes the feature vector of node $i$, while $\mathbf{W}$ and $\mathbf{a}$ represent the learnable parameters. In our method, the adjacency matrix obtained using lsAGC, $\mathbf{F} = [f_{\mathbf{x}_s \to -\mathbf{x}_t}]$, is considered equal to the node features $\mathbf{H} = [\mathbf{h}_i]_{i=1}^{N}$, with $\mathbf{F} = \mathbf{H}$, signifying that the graph's structure has an immediate impact on the feature representations.. The feature matrix $\mathbf{H} \in \mathbb{R}^{N \times d}$ encapsulates the nodes' attributes, where $N$ is the number of nodes, and $d$ is the dimensionality of the feature space.

The GAT operates on these representations to model intricate relationships between nodes. Attention scores are computed based on the adjacency and feature matrices, and are only calculated for connected nodes, i.e., where $F_{ij} = 1$.

### 3.2.3 Multi-Head Attention and Aggregation

The Graph Attention Networks (GATs) often utilize a multi-head attention mechanism, which applies multiple sets of learnable parameters to capture diverse facets of relationships between nodes. This method enhances stability in learning and enriches the model's expressiveness by integrating information across various attention heads. In our approach, we averaged the outputs from different heads instead of concatenating them, as given by:

$$\mathbf{h}_i = \text{Aggregate}(\alpha_{ij} \cdot \mathbf{W}\mathbf{h}_j), \forall j \in N(i) \tag{7}$$

The proposed model leverages the adjacency matrix to guide attention, focusing on pertinent relationships while considering connected nodes' features.

### 3.2.4 Loss Function

In our approach, we utilize the Binary Cross-Entropy Loss with Logits ($L_{BCE}$) as our chosen loss function. This loss function is applied to the outputs of a linear layer, which represents the predicted class logits, and the encoded feature representations. The modified loss function is formulated as follows:

$$L_{loss} = L_{BCE}(y, \hat{y}) + L_{BCE}(y_{enc}, y) \tag{8}$$

Here, $\hat{y}$ signifies the predicted class logits stemming from the linear layer, $y_{\text{enc}}$ denotes a transformed encoded representation, and $y$ corresponds to the vector containing the target labels. This loss formulation is operationalized during both the training and validation phases. By combining these two components within the loss, we aim to emphasize distinct facets of the learning task. This approach holds potential for heightened performance and improved generalization across the model's predictions.

### 3.3  Data Augmentation

Data augmentation is essential for enhancing the diversity of training datasets, mitigating overfitting, and improving generalization in deep learning models. In our study, we exploited specific parameters of the lsAGC method, namely the retained components of PCA and the autoregressive model number, to generate diverse adjacency matrices for each dataset. This augmentation strategy was integrated into the training process, markedly increasing the training sample size. We employed a data augmentation factor of 27, expanding an initial dataset of 60 subjects to a total of 1620 samples. This expansion was instrumental in constructing a robust dataset, a critical factor for the effective training of the Graph Attention Network (GAT) model in our research.

## 4  RESULTS

Mean connectivity matrices that were extracted using lsAGC and cross-correlation, are shown in Fig. 2 for both typical control and frequent marijuana user cohorts. Different patterns are visible to the naked eye for both methods. In the following, we quantitatively investigate the difference between connectivity patterns of the two subject cohorts using an MVPA approach.

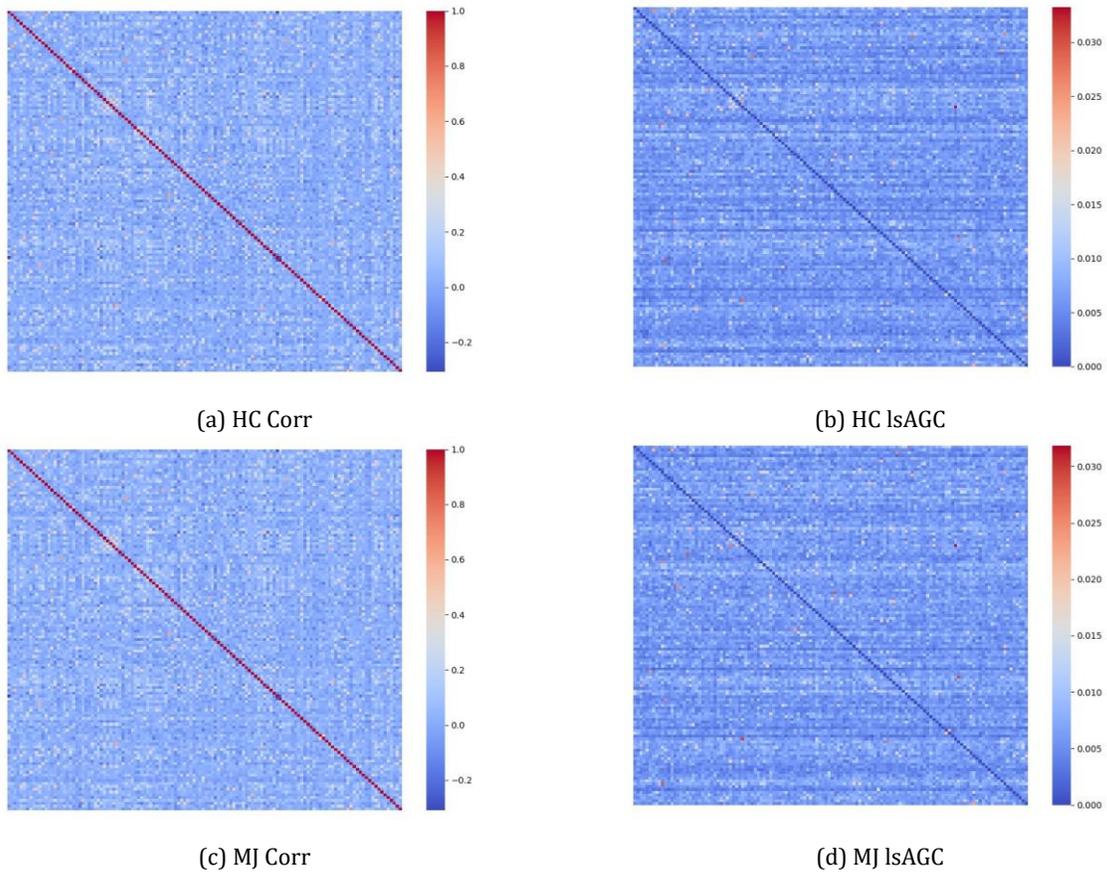

Figure 2: Mean connectivity matrices of control subjects and marijuana users using different methods. Variations in connectivity patterns are observed between the groups.

Table 2: Accuracy Results Comparison for Correlation Coefficient, Random Guess, and Our Method (lsAGC) for five-fold test measures on the dataset.

|   | lsAGC Accuracy (%) | Correlation Accuracy (%) | Random Guess Accuracy (%) | Std |
| --- | --- | --- | --- | --- |
| 0 | 62.28 | 53.12 | 46.15 | |
| 1 | 62.53 | 53.32 | 56.03 | |
| 2 | 59.73 | 55.92 | 45.06 | |
| 3 | 61.65 | 52.93 | 48.00 | |
| 4 | 63.18 | 49.61 | 40.00 | |

| | | | |
|---|---|---|---|
| Mean | **61.47** | 52.98 | 47.05 |
| Std | **1.44** | 1.65 | 6.25 |

The results presented in Table 2 provide insights into the performance of the correlation coefficient method, random guess method, and our method, lsAGC, across five distinct folds. The accuracy values for each method are listed in separate columns, and the results for all three methods are compared across the same five folds.

The mean accuracy computed for the correlation coefficient is approximately 52.98%, while the mean accuracy for our method, lsAGC, is approximately 61.47%. The mean accuracy for random guess method is approximately 47.05%. The standard deviation (Std) for the correlation coefficient method is approximately 1.65, and for the random guess method, it is approximately 6.25. The standard deviation for our method (lsAGC) is approximately 1.44.

These combined results collectively offer valuable insights into the efficacy of the correlation coefficient, random guess, and our method for the specific task. Despite observable performance variance across different folds, the overall performance indicates that our method demonstrates higher accuracy within the experimental setup.

## 5 CONCLUSIONS

To sum up, the use of large-scale Augmented Granger Causality (lsAGC) as a possible marker to differentiate marijuana users from regular controls is explored in our research by employing restingstate functional Magnetic Resonance Imaging (fMRI). Changes in brain network connectivity connected to marijuana consumption are built upon prior studies, reassessed by probing the ability of lsAGC to effectively distinguish these modifications. The description of directed causal connections within the complex fluctuations of fMRI time-series information is facilitated by lsAGC through a multivariate technique focusing on dimensionality reduction, augmenting data, and time-series predictive modeling. Brain connections are identified using the lsAGC process and used as vital elements for our categorization efforts in a dataset that includes 60 adult subjects who were diagnosed with childhood ADHD from the Addiction Connectome Preprocessed Initiative (ACPI) repository. A

Graph Attention Neural Network (GAT) is used for the categorization task, selected for its competence in handling graph-oriented data and identifying complex connections among various brain areas. Consequently, new possibilities for employing lsAGC-based brain network connectivity to separate marijuana users from standard controls are uncovered by our research, highlighting its potential to enhance our comprehension of the neurological effects tied to marijuana consumption. Further scrutiny and validation through upcoming prospective clinical experiments will be essential to confirm the clinical relevance of our method.

## 6 Acknowledgements

This research was partially funded by the American College of Radiology (ACR) Innovation Award "AI-PROBE: A Novel Prospective Randomized Clinical Trial Approach for Investigating the Clinical Usefulness of Artificial Intelligence in Radiology" (PI: Axel Wismüller) and an Ernest J. Del Monte Institute for Neuroscience Award from the Harry T. Mangurian Jr. Foundation (PI: Axel Wismüller). This work was conducted as a Practice Quality Improvement (PQI) project related to the American Board of Radiology (ABR) Maintenance of Certificate (MOC) for A.W.